\documentclass[prl,twocolumn,notitlepage,reprint,amsfonts,amssymb,amsmath,superscriptaddress,floatfix,showpacs,longbibliography]{revtex4-1}

\usepackage{graphicx}
\usepackage[colorlinks]{hyperref}
\hypersetup{linkcolor=blue,citecolor=blue,urlcolor=blue}

\newcommand{\imperial}{
Blackett Laboratory,
Department of Physics,
Imperial College London,
London SW7~2AZ,
United Kingdom
}
\newcommand{\UAM}{
Departamento de F\'{i}sica Te\'{o}rica
de la Materia Condensada and Condensed Matter Physics Center (IFIMAC),
Universidad Aut\'{o}noma de Madrid,
E-28049 Madrid,
Spain
}

\newcommand{\DIPC}{
Donostia International Physics Center (DIPC),
 E-20018 Donostia/San Sebastian,
 Spain
}

\begin{document}

\title{Nonequilibrium plasmon emission drives ultrafast carrier relaxation
dynamics in photoexcited graphene}

\author{J.~M.~Hamm}
\email{j.hamm@imperial.ac.uk}
\affiliation{\imperial}
\author{A.~F.~Page}
\affiliation{\imperial}
\author{J.~Bravo-Abad}
\affiliation{\UAM}
\author{F.~J.~Garcia-Vidal}
\affiliation{\UAM}
\affiliation{\DIPC}
\author{O.~Hess}
\email{o.hess@imperial.ac.uk}
\affiliation{\imperial}

\date{\today}

\pacs{73.20.Mf, 78.67.Wj, 71.10.Ca, 72.10.Di}

\begin{abstract}
The fast decay of carrier inversion in photoexcited graphene has been attributed
to optical phonon emission and Auger recombination.
Plasmon emission provides another pathway that, as we show here, drives the
carrier relaxation dynamics on ultrafast timescales.
In studying the nonequilibrium relaxation dynamics we find that plasmon emission
effectively converts inversion into hot carriers, whose energy is then extracted
by optical phonon emission.
This mechanism not only explains the observed fs-lifetime of inversion but also
offers the prospect for atomically thin ultrafast plasmon emitters.
\end{abstract}

\maketitle

Graphene owes its extraordinary electronic, optical and plasmonic properties
\cite{CastroNeto2009,Koppens2011,Bonaccorso2010,Popov2012,Stauber2014}
to the presence of Dirac points in its band structure, in the vicinity
of which electrons are described as massless Dirac fermions (MDFs).
One of graphene's unique characteristics is that low-energy pair excitations
can decay into plasmons \cite{Giuliani1982,Bostwick2007}, a process that is
kinematically forbidden in most 3D and 2D materials.

When excited with a femtosecond optical pulse, a hot
carrier plasma is generated, which rapidly thermalizes into a state
of quasi-equilibrium due to carrier-carrier scattering on 10-fs scales
\cite{Ryzhii2007,George2008}. After thermalization, electrons and holes recombine
via interband scattering processes, effectively depleting carrier
inversion over time \cite{Malic2011,Sun2012}. Femtosecond pump-probe experiments
\cite{Breusing2009,Li2012,Brida2013}
and time- and angle-resolved photo-emission spectroscopy (tr-ARPES)
\cite{Gierz2013, Johannsen2013,Gierz2015}
indicate the presence of two dominant recombination channels: a slow decay on a
1-ps scale due to emission of optical phonons
\cite{Butscher2007,Rana2009,Wang2010},
whose signatures can be observed by Raman spectroscopy, and a fast
decay on a 100-fs scale that has been attributed to Auger recombination (AR)
\cite{Breusing2011,Brida2013}.
This interpretation remains subject to discussion as AR processes are supressed
when considering dynamic screening in random-phase approximation (RPA)
\cite{Tomadin2013,Tielrooij2013}.

\begin{figure}[t]
\includegraphics{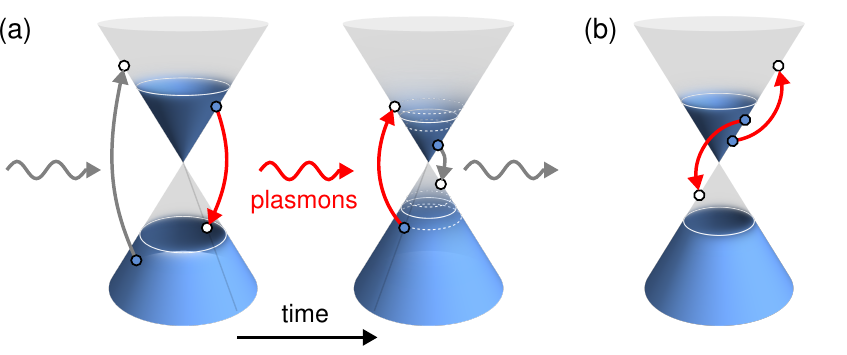}
\caption{\label{fig:TitleFig}(Color online) Schematic of carrier recombination
channels in inverted graphene. (a) Nonequilibrium plasmon emission
and (b) Auger recombination both convert electron/hole pairs into
hot carriers. In contrast to Auger recombination, plasmon emission
is an inelastic process that is enabled by the strong coupling of
electron/hole pair to 2D plasmon excitations. }
\end{figure}

In this Letter we show that nonequilibrium plasmon emission (NPE)
(see Fig.~\ref{fig:TitleFig}) provides an alternative pathway for
carrier relaxation that drives the decay of inversion on a 100-fs
scale. It was recently established that plasmon emission in inverted
graphene is ultrafast, with rates that exceed those of optical phonon
emission by at least one order of magnitude \cite{Rana2011,Page2015}.
However, in assuming predefined carrier and plasmon distributions,
these works did not account for the dynamic evolution of the coupled
carrier-plasmon system far from equilibrium. Thus, despite the prediction
of ultrafast rates, it remains an open question how NPE impacts
the relaxation dynamics, as plasmons are not only constantly emitted
but also absorbed back into the electron/hole plasma. Ultimately,
to assess the importance of NPE as channel for carrier relaxation,
it is necessary to study the temporal evolution of the nonequilibrium
plasmon distribution in interaction with the inverted carrier system,
and together with optical phonon emission, which provides additional
pathways for carrier recombination and cooling. As we will show, an
interplay of channels emerges, where NPE drives the ultrafast conversion
of inversion into hot carriers on 100-fs scales, while optical phonons
efficiently extract heat from the MDF plasma.

As basis for this work we introduce nonequilibrium rate equations
for an inverted MDF plasma in interaction with bosonic reservoirs
via emission and absorption processes. In quasi-equilibrium, the state
of the inverted carrier plasma is characterized by chemical potentials
$\mu^{\alpha}$ ($\alpha=e,h$) for the electrons and holes and a
common plasma temperature $T_c$, whose temporal changes can be expressed
by those of the carrier densities $N^{\alpha}=N(\mu^{\alpha},T_c)$
of electron and holes and the total energy density $U=U^{e}+U^{h}$
of the plasma {[}where $U^{\alpha}=U(\mu^{\alpha},T_c)${]}
(see \cite{supp}, Sec.~I).
We split the time-derivatives into optical excitation (pump) and relaxation
terms,
\begin{equation}
\dot{N}^{\alpha}=\dot{N}|_{\mathrm{pump}}-\dot{N}|_{\mathrm{rel}}\:,\quad\text{and}\quad\dot{U}=\dot{U}|_{\mathrm{pump}}-\dot{U}|_{\mathrm{rel}}\;,
\end{equation}
where $\dot{N}^{e}=\dot{N}^{h}$ as particles and holes are created
(and annihilated) pairwise. For quasi-mono\-chro\-ma\-tic excitation
with frequency $\hbar\omega$ and intensity $I_{\omega}(t)$ the pump
terms are given by $\dot{N}|_{\mathrm{pump}}=\mathrm{Re}[\sigma_{\mathrm{inter}}(\omega)]|E_{\omega}(t)|^{2}/(2\hbar\omega)$
and $\dot{U}|_{\mathrm{pump}}=\mathrm{Re}[\sigma(\omega)]|E_{\omega}(t)|^{2}$/2.
Here $Z_0$ is the vacuum impedance, $|E_{\omega}(t)|^{2}=2Z_{0}|t(\omega)|^{2}I_{\omega}(t)$ 
the electric field square, $t(\omega)=(1+Z_{0}\sigma(\omega)/2)^{-1}$
the transmission coefficient, $\sigma=\sigma(\omega;\mu^{e},\mu^{h},T_c)$
the optical conductivity, and $\sigma_{\mathrm{inter}}$ its interband
only contribution.

Carrier relaxation, in turn, is a result of the interaction with bosonic
reservoirs, such as the plasmon and optical phonon fields
\cite{Pines1962}.
The microscopic mechanisms behind the dynamic relaxation of the carrier number
and energy densities are intra- and interband scattering processes, as
given by Boltzmann collision integrals. To account for the various bosonic
reservoirs ($\nu$), as well as intra- and intraband transitions ($\lambda$),
we write
\begin{equation}
\dot{N}|_{\mathrm{rel}}=\sum_{\nu}R_{\nu,eh}\:,\quad\dot{U}|_{\mathrm{rel}}=\sum_{\nu}\sum_{\lambda=ee,hh,eh}S_{\nu,\lambda}\:,\label{eq:RelDyn}
\end{equation}
The net carrier and energy relaxation rates $R_{\nu,eh}$ and $S_{\nu,\lambda}$
are obtained by summation over all wavevector states $\mathbf{q}$
of the reservoir bosons, i.e.,
\begin{equation}
R_{\nu,\lambda}=\frac{1}{A}\sum_{\mathbf{q}}\, r_{\nu,\lambda}(\mathbf{q})\:,\quad S_{\nu,\lambda}=\frac{1}{A}\sum_{\mathbf{q}}\epsilon_{\nu}(\mathbf{q})r_{\nu,\lambda}(\mathbf{q})\:,\label{eq:RelaxRates}
\end{equation}
where
\begin{equation}
r_{\nu,\lambda}(\mathbf{q})=\gamma_{\nu,\lambda}^{+}(\mathbf{q})[n_{\nu}(\mathbf{q})+1]-\gamma_{\nu,\lambda}^{-}(\mathbf{q})n_{\nu}(\mathbf{q})\label{eq:SpecRate}
\end{equation}
is the net spectral emission rate, \emph{$A$} the area of the graphene
sample, $\epsilon_{\nu}(\mathbf{q})=\hbar\omega_{\nu}(\mathbf{q})$
the energy-dispersion of the boson field, $\gamma_{\nu,\lambda}^{+}$
($\gamma_{\nu,\lambda}^{-}$) the emission (absorption) rate, and
$n_{\nu}(\mathbf{q})$ the nonequilibrium distribution of the respective
boson field. The latter evolves in time according to
\begin{equation}
\begin{split}\dot{n}_{\nu}(\mathbf{q})=\sum_{\lambda=ee,hh,eh} & r_{\nu,\lambda}(\mathbf{q})-\left.\dot{n}_{\nu}(\mathbf{q})\right|_{\mathrm{decay}}\end{split}
\label{eq:OccEq}
\end{equation}
with a decay term that accounts for additional loss channels.
Within the relaxation time approximation we define
$\left.\dot{n}_{\nu}(\mathbf{q})\right|_{\mathrm{decay}}=\tau_{\nu}^{-1}[n_{\nu}(\mathbf{q})-n_{\nu}^{(\mathrm{eq})}(\epsilon_{\nu}(\mathbf{q}),T_{0})]$,
introducing the decay rate
$\tau_{\nu}^{-1}$,
and the equilibrium distribution
$n_{\nu}^{(\mathrm{eq})}(\epsilon,T_{0})$
of the reservoir at ambient temperature $T_{0}$.
Note that, as
$n_{\nu}(\mathbf{q})$
returns to equilibrium,
$r_{\nu,\lambda}(\mathbf{q})$
becomes zero according to the condition of detailed balance.

Equipped with these equations, we proceed to study the carrier relaxation
dynamics for interaction with the plasmon reservoir and then under
the inclusion of phonon channels. The results in this work are obtained
by integrating Eqs.~(\ref{eq:RelDyn}) and (\ref{eq:OccEq}) over
time. For all our simulations we consider a suspended sheet of graphene
at ambient temperature that is excitated by a pulse with a fluence
of $133\,\mu\mathrm{J}/\mathrm{cm^{2}}$ and a photon energy of $1\,\mathrm{eV}$.
To enforce the same initial conditions for all simulations, we keep all
relaxation channels switched off during optical excitation.
Under these conditions the excitation pulse creates a thermalized inverted
state with chemical potentials $\mu^{e}=\mu^{h}\approx0.3\,\mathrm{eV}$
and a carrier temperature $T_c\approx2288\,\mathrm{K}$.

\begin{figure}
\includegraphics{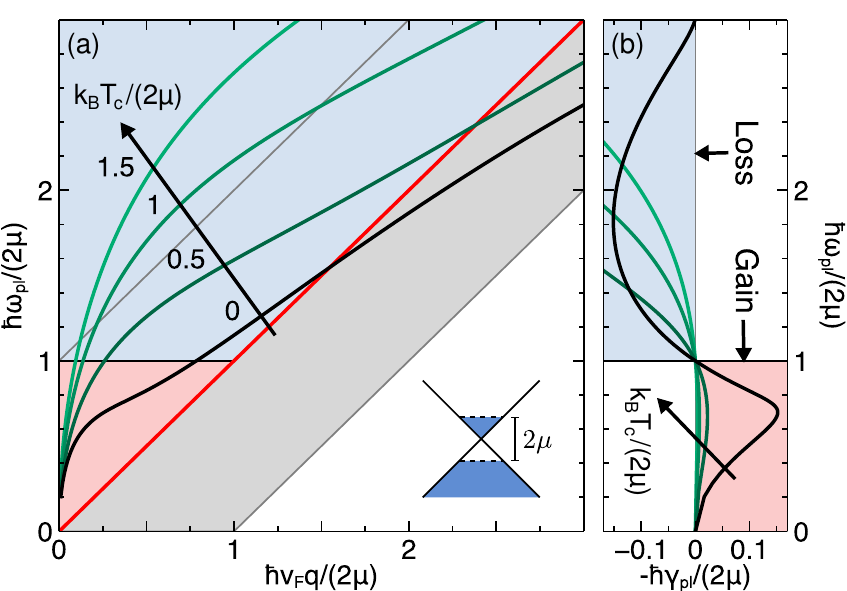}
\caption{\label{fig:PlDispAndRates}(Color online)
(a) Plasmon frequency dispersion $\omega_{\mathrm{pl}}(q)$ and
(b) loss spectrum $\gamma_{\mathrm{pl}}(\omega)$ of intrinsic graphene for
varying temperature $T_c$ scaled with chemical potential $\mu$.
The dispersion crosses through regimes of interband gain (red area) and
absorption (blue area).
For $\mu=0.1\:\mathrm{eV}$ temperatures shown range from
$T_c=0-3208\:\mathrm{K}$.}
\end{figure}

When the carrier system of graphene is inverted, electron/hole pair
excitations and plasmons interact strongly via plasmon emission and
absorption processes. The associated spectral rates can be calculated
approximatively using Fermi golden rule (FGR), with an accuracy that critically
depends on the exactness of the plasmon frequency dispersion
$\omega_{\mathrm{pl}}(q)$ \cite{Page2015}.
Therefore we trace the\emph{ }exact\emph{ complex-frequency} roots
$\omega(q)=\omega_{\mathrm{pl}}(q)-i\gamma_{\mathrm{pl}}(q)$ of the
dynamic dielectric function, i.e., the solutions of 
\begin{equation}
\varepsilon_{\mathrm{RPA}}(q,\omega)=1-V_{q}\Pi|_{\mu^{e},\mu^{h}}^{T_c}(q,\omega)=0.\label{eq:PlasmonDisp}
\end{equation}
Here $V_{q}=e^{2}/(2\varepsilon_{0}q)$ is the bare 2D
Coulomb potential and $\Pi|_{\mu^{e},\mu^{h}}^{T_c}(q,\omega)$ the
irreducible nonequilibrium polarizability of inverted graphene at
temperature $T_c$.
Figure~\ref{fig:PlDispAndRates} shows frequency dispersion
$\omega_{\mathrm{pl}}(q)$ and the loss spectrum $\gamma_{\mathrm{pl}}(\omega)=\gamma_{\mathrm{pl}}(q_{\mathrm{pl}}(\omega))$
for temperatures $k_{B}T_c/(2\mu)$ in the range of $0-1.5$. Plasmons
experience interband gain for $\hbar\omega_{\mathrm{pl}}<2\mu$ (red
area) and inter- and intraband loss for $\hbar\omega_{\mathrm{pl}}>2\mu$
(blue and grey areas) at all temperatures, as evident from the change of sign of
the net stimulated absorption rate
$\gamma_{\mathrm{pl}}^{\mathrm{stim}}(\omega)=2\gamma_{\mathrm{pl}}(\omega)$.
Note, for our analysis, it is sufficient to consider interband processes
only, as reabsorption above the Fermi-edge rapidly depletes the plasmon
mode population before entering the intraband regime.
The respective interband emission/absorption rates are given by
\begin{equation}
\begin{split}\gamma_{\mathrm{pl},eh}^{\pm}(q)\approx & \frac{2\alpha_{g}\theta(\omega-v_{F}q)}{\sqrt{\left(\frac{\omega}{v_{F}q}\right)^{2}-1}}\frac{K_{eh}^{\pm}|_{\mu^{e},\mu^{h}}^{T_c}\left(q,\omega\right)}{\frac{\partial\mathrm{Re}[\varepsilon_{\mathrm{RPA}}(q,\omega)]}{\partial\omega}}\Biggr|_{\omega=\omega_{\mathrm{pl}}(q)}.\end{split}
\label{eq:InterPlRate}
\end{equation}
Here $\alpha_{g}=\alpha_{f}c/v_{F}\approx300/137$ is the fine-structure
constant of graphene and $K_{eh}^{\pm}|_{\mu^{e},\mu^{h}}^{T_c}(q,\omega)$
a dimensionless measure for the phase-space of the absorption and
emission processes
(see \cite{supp}, Sec.~II).
In addition to these processes,
one needs to consider the loss of plasmons due to collisions \cite{Low2014},
which emerges as a result of various elastic and inelastic velocity scattering
processes \cite{Lin2014}.
Values for $\tau_{\mathrm{coll}}$ are typically in the 50-500 fs range,
depending on carrier temperature, chemical potentials and impurity
concentration.
While collision loss only weakly impacts on the frequency dispersion
$\omega_{\mathrm{pl}}(q)$ \cite{Page2015}, it serves as a secondary decay
channel as plasmons purged from the reservoir are not available for reabsorption at
later times.

\begin{figure}
\includegraphics{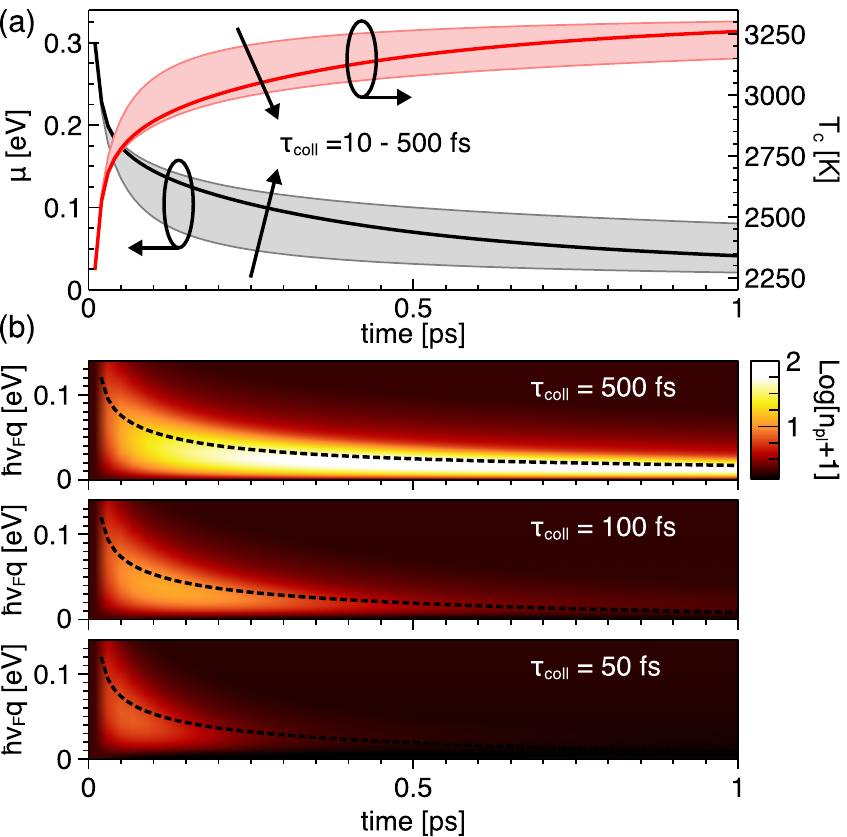}
\caption{\label{fig:PlRelaxDyn}(Color online) Relaxation dynamics of the coupled
electron/hole and plasmon system. (a) Chemical potential $\mu$ (black)
and carrier temperature $T_c$ (red) for $\tau_{\mathrm{coll}}=100\,\mathrm{fs}$
(solid lines) and varying collision time (shaded area). (b) Temporal
evolution of nonequilibrium plasmon distributions $n_{\mathrm{pl}}(q)$
and emission/absorption edge (black dashed lines).}
\end{figure}

We first study the interaction of carriers with plasmons in isolation,
i.e., without inclusion of the optical phonon channels.
Figure~\ref{fig:PlRelaxDyn} shows the evolution of the chemical potential
($\mu=\mu^{e}=\mu^{h}$) and the carrier temperature $T_c$ for varying collision
times, together with the temporal traces of the nonequilibrium plasmon
distribution $n_{\mathrm{pl}}(q)$.
Within the first $100\,\mathrm{fs}$ after photoexcitation, a burst of plasmon
emission occurs that leads to a drop of inversion (solid black line) to roughly
half its initial value;
at the same time, the carrier temperature (solid red line) increases by around
$500\,\mathrm{K}$ {[}see Fig.~\ref{fig:PlRelaxDyn}(a){]} as plasmons are emitted
below the Fermi-edge.
Over the next $200\,\mathrm{fs}$ the recombination process gradually slows down.
This is because reabsorption of plasmons above the Fermi-edge and the decrease
of the plasmon emission rate with temperature creates a
\emph{plasmon emission bottleneck}.
The temporal traces of the nonequilibrium plasmon distribution
$n_{\mathrm{pl}}(q,t)$ {[}Fig.~\ref{fig:PlRelaxDyn}(b){]} are shown together
with the threshold wavevector $q_{\mathrm{th}}(t)$ (black dashed line) for which
the net plasmon gain is zero, i.e., $\gamma_{\mathrm{pl}}(q_{\mathrm{th}})=0$
(no plasmon gain/loss).
Below this threshold plasmons are emitted, above they are reabsorbed.
Collision loss partially removes the emission bottleneck as it purges plasmons
from the reservoir and thus prevents reabsorption of plasmons into the
electron/hole plasma.
As a result, the recombination of carriers accelerates with increasing collision
rate {[}grey shaded area in Fig.~\ref{fig:PlRelaxDyn}(a){]}. 

Having analyzed the plasmon channel in isolation, we next study the interplay
of NPE and optical phonon emission, which plays a pivotal role
as they facilitate the recombination of electron/hole pairs \cite{Rana2009},
the cooling of hot carriers \cite{Wang2010}, as well as collision
loss \cite{DasSarma2011}. In inverted graphene, or at high temperature,
the inter- and intraband emission of longitudinal and transverse optical
(LO/TO) phonons are dominant channels for carrier recombination and
cooling. For this work we consider all relevant optical phonon channels,
the $\Gamma\mathrm{O}$, the $\mathrm{KO},$ and the $\mathrm{KA}$
phonons with energies of $\epsilon_{\mathrm{\Gamma O}}=196\,\mathrm{meV}$,
$\epsilon_{\mathrm{KO}}=160\,\mathrm{meV}$, and
$\epsilon_{\mathrm{KA}}=120\,\mathrm{meV}$, respectively
\cite{Borysenko2010,Fang2011}.
As the optical phonon modes are quasi dispersion-free, Eq.~(\ref{eq:RelaxRates})
reduces to
$R_{\nu,\lambda}=\Gamma_{\nu,\lambda}^{+}[n_{\nu}+1]-\Gamma_{\nu,\lambda}^{-}n_{\nu}$
and $S_{\nu,\lambda}=\epsilon_{\nu}R_{\nu,\lambda}$;
where $n_{\nu}$ is the occupation number and $\epsilon_{\nu}$ the phonon
energy.
Closed-form expressions for the rates
$\Gamma_{\nu,\lambda}^{\pm}=M_{\nu,\lambda}\gamma_{\nu,\lambda}^{\pm}$
and the phonon density of states $M_{\nu,\lambda}$ are given in the supplement
\cite{supp}, Sec.~III.

\begin{figure*}
\includegraphics{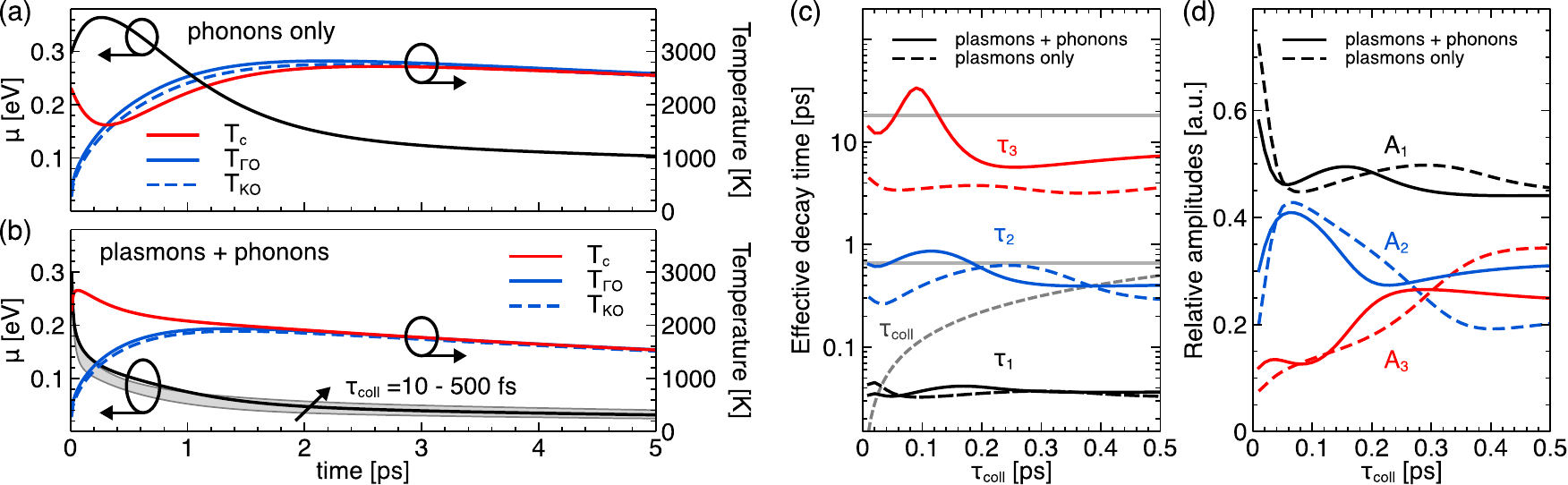}

\caption{\label{fig:PlPhoRelaxDyn}(Color online)
Relaxation dynamics of carriers coupled to (a) phonon only;
(b) plasmons and phonons where $\tau_{\mathrm{coll}}=100\,\mathrm{fs}$.
Shown are chemical potential $\mu$ (black line),
carrier temperature $T_c$ (red line), and temperatures of the
$\Gamma\mathrm{O}$, $\mathrm{KO}$ phonons $T_\mathrm{\Gamma O}$ (blue solid line),
and $T_\mathrm{K O}$ (blue dashed line);
the grey area in (b) indicates change with collision time.
(c)+(d) Effective decay times $\tau_i$ and amplitudes $A_i$ extracted from
tri-exponential fit to $\mu(t)$ considering plasmons only (dashed lines),
plasmons and phonons (solid lines); grey lines indicate decay times extracted
from (a) (phonons only).}
\end{figure*}
To understand the fundamental difference of NPE and optical phonon
emission, we first consider the carrier dynamics without NPE, i.e.,
under inclusion of optical phonon emission only assuming a phonon
decay time of $\tau_{\mathrm{lat}}=2.5\:\mathrm{ps}$ \cite{Wang2010}.
Figure~\ref{fig:PlPhoRelaxDyn}(a) depicts the temporal evolution of
the chemical potential (black line) together with the carrier temperature
(red line) and temperatures of the $\Gamma\mathrm{O}$, $\mathrm{KO}$
phonons (blue lines; $\mathrm{KA}$ phonons omitted for clarity).
The initial dip in carrier temperature (at $t\approx400\,\mathrm{fs}$),
accompanied by a rise of phonon temperatures and chemical potential,
is predominantly due to intraband phonon emission, which continuously
extracts energy from the carrier plasma. As the phonon temperature
equilibrates with the plasma temperature, intraband cooling becomes
increasingly inefficient. Over the next 1-ps, $\mu$ drops from
$0.3\,\mathrm{eV}$ to $0.2\,\mathrm{eV}$ mainly due to interband emission of
optical phonons.
As carrier temperature and chemical potential further decrease both inter- and
intraband emission of optical phonons slows down.
Less electron/hole pairs are available at the required phonon energies and
carrier cooling is bottlenecked by the slow decay of optical phonons into
acoustic phonons on ps-scales.

The relaxation dynamics changes dramatically when NPE is taken into
account {[}see Fig.~\ref{fig:PlPhoRelaxDyn}(b){]}.
The rapid drop of inversion within the first 100-fs, due to the plasmon emission
burst, is followed by a gradual slowdown as plasmons are reabsorbed above
the Fermi-edge.
The plasmon energy that flows back into the electron/hole system heats the
carrier plasma and thus prevents the drop in carrier temperature that was
observed in Fig.~\ref{fig:PlPhoRelaxDyn}(a).
At $t=1\,\mathrm{ps}$ the chemical potential has fallen well below
$0.1\,\mathrm{eV}$, compared to a value of $\mu\approx0.25\,\mathrm{eV}$
in Fig.~\ref{fig:PlPhoRelaxDyn}(a), where plasmon emission was switched off.
Collision loss further accelerates the decay of inversion, albeit not as
strongly as in Fig.~\ref{fig:PlRelaxDyn}(a) (plasmons only), as phonons now
provide an efficient cooling channel that alleviates the impact of carrier
heating due to plasmon reabsorption.
The combination of plasmon and optical phonon emission effectively bypasses
bottlenecks observed for the isolated channels, thereby accelerating the carrier
recombination and cooling dynamics.

To analyze the decay of carrier inversion due to NPE we fit $\mu(t)$ with
tri-exponential functions
$\mu_\mathrm{fit}(t)=\mu_{0}\sum_{i=1}^{3}A_{i}\exp(-t/\tau_{i})$.
The extracted (effective) decay times $\tau_{i}$ and relative
amplitudes $A_{i}$ are shown in
Fig.~\ref{fig:PlPhoRelaxDyn}(c, d) in dependence on collision time.
NPE alone induces three timescales (dashed lines):
a fast rate ($\tau_{1}\approx30\,\mathrm{fs}$) that relates to the plasmon
emission into unoccupied plasmon modes with an amplitude that rises quickly for
$\tau_{\mathrm{coll}}<\tau_{1}$;
a slower rate ($\tau_{1}\approx300\,\mathrm{fs}$) due to fast emission and
reabsorption at the Fermi-edge;
and a slow rate ($\tau_{3}\approx3\,\mathrm{ps}$) that is strongly influenced by
absorption below the Fermi-edge, as evident from the rise of amplitude $A_{3}$
with $\tau_{\mathrm{coll}}$.
Activating the phonon channel in addition to NPE (solid lines) causes an
interplay of channels as the NPE decay times $\tau_{2}$ and $\tau_{3}$ are
close to the timescales that govern the decay via phonon emission
{[}Fig.~\ref{fig:PlPhoRelaxDyn}(c); grey lines{]}.
This is apparent in the change of amplitudes $A_{2}$ and $A_{3}$, which for
$\tau_{\mathrm{coll}}>200\,\mathrm{fs}$ become almost constant, resulting in a
decay of inversion that is almost independent of $\tau_{\mathrm{coll}}$
{[}see Fig.~\ref{fig:PlPhoRelaxDyn}(b); grey shaded area{]}. Most prominently, the relaxation dynamics and extracted timescales show that inversion above $\mu_e = \mu_h = 0.1\,\text{eV}$ decays on 100-fs scales due to the ultrafast decay of electron/hole pairs into plasmons.

In conclusion, we have established that plasmon emission drives the ultrafast carrier relaxation in photo-excited graphene, as 2D plasmons can couple strongly to pair excitations of the inverted carrier plasma. Our results are consistent with the recent experimental observation of fs-decay of population inversion \cite{Li2012,Gierz2013}. In contrast to Auger processes, whose experimental detection is very challenging \cite{Gierz2015}, plasmon emission can be directly observed and thus provides a novel path for experimental characterization of relaxation processes. Interaction of plasmons with a (polar) substrate causes a red-shift of the plasmon emission spectrum and hybridization with surface optical phonons. These effects are not expected to change the main conclusions drawn here, and will be studied in a future work.

The authors thank Fouad Ballout for discussions.
This work has been funded by the Engineering and Physical Sciences
Research Council (United Kingdom), the Leverhulme Trust (United Kingdom),
the European Research Council (ERC-2011-AdG Proposal No. 290981) and
the Spanish MINECO (MAT2011-28581-C02-01 grant).


%

\end{document}